\documentclass[journal]{IEEEtran}
\usepackage[T1]{fontenc}  
\usepackage[english]{babel}
\usepackage{graphicx}
\usepackage{subcaption}
\usepackage{multirow}
\usepackage{amsmath}
\usepackage{upgreek}
\usepackage{hyperref}
\usepackage{epstopdf}

\newcommand{\cv}{C_\mathrm{V}}
\newcommand{\testfun}{v_{ip}^{(k)}}
\newcommand{\shapefun}{v_{jq}^{(k)}}
\newcommand{\dd}{\mathrm{d}}
\newcommand{\lobatto}{\mathrm{LO}}

\begin{document}
\title{Quasi-3D Thermal Simulation of Quench Propagation in Superconducting Magnets}%
\author{L.~A.~M.~D'Angelo, J.~Christ, and H.~De Gersem%
\thanks{L.~A.~M.~D'Angelo is with the Institut f\"ur Teilchenbeschleunigung und Elektromagnetische Felder (TEMF) and with the Centre for Computational Engineering, Technische Universit\"at Darmstadt, Darmstadt 64289, Germany (e-mail: dangelo@temf.tu-darmstadt.de).}%
\thanks{J.~Christ is with the Institut f\"ur Teilchenbeschleunigung und Elektromagnetische Felder (TEMF), Technische Universit\"at Darmstadt, Darmstadt 64289, Germany (e-mail: jonas.christ@stud.tu-darmstadt.de).}%
\thanks{H.~De Gersem is with the Institut f\"ur Teilchenbeschleunigung und Elektromagnetische Felder (TEMF) and with the Centre for Computational Engineering, Technische Universit\"at Darmstadt, Darmstadt 64289, Germany (e-mail: degersem@temf.tu-darmstadt.de).} }%
\maketitle 

\begin{abstract}
To deal with the multi-scale nature of the quench propagation problem in superconducting magnets, this work presents a quasi-three-dimensional (Q3D) approach combining a two-dimensional finite-element method (FEM) in the transversal cross-section of the magnet for resolving the geometrical details, with a one-dimensional spectral-element method based on orthogonal polynomials in longitudinal direction for accurately and efficiently representing the quench phenomena. The Q3D formulation is elaborated and the idea is illustrated on a thermal benchmark problem. Finally, the method is validated against a conventional 3D simulation carried out by a commercial software. In terms of computational efficiency, it is shown that the proposed Q3D approach is superior to the conventional 3D FEM.
\end{abstract}	


\section{Introduction}
\IEEEPARstart{S}{uperconducting} accelerator magnets are used in the Large Hadron Collider (LHC) at CERN to achieve high magnetic fields while keeping the energy consumption low~\cite{Russenschuck_2010aa}. This comes at the cost of having to deal with \emph{quenches}~\cite{Wilson_1983aa}, which can lead to extensive damages of the magnet, e.g.~reported in~\cite{Bajko_2009aa}, making a reliable quench protection system mandatory. Quench protection systems are triggered upon quench detection, i.e.~when predefined voltage or resistance thresholds are exceeded~\cite{Steckert_2019aa}. However, today, these thresholds are equipped with a large safety margin resulting into an overly sensitive quench detection system leading to frequent and unnecessary shutdowns, which considerably reduce the availability of the whole LHC machine due to false triggers~\cite{Apollonio_2015aa}. This issue becomes even more critical for the newly developed magnets based on the Nb$_3$Sn technology~\cite{Ambrosio_2015aa}.

\begin{figure}[tbp]
	\centering
	\includegraphics[width=1\columnwidth]{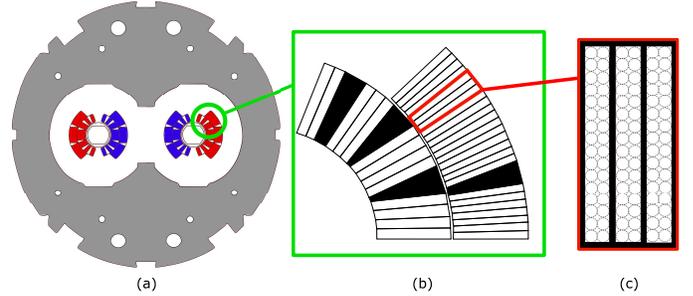}
	\caption{In (a), the cross-section of the main dipole magnet in the LHC is shown. The coils are depicted in blue and red depending on the current direction, the iron yoke is colored in gray. In (b), the shell-type configuration of the coils is illustrated in detail showing the rectangular Rutherford cables and the trapezoidal copper keystones. The benchmark model (c) represents a stack of three Rutherford cables (white) wrapped by glass fibre insulation (black).}
	\label{fig:benchmark}
\end{figure}

Numerical simulations play a crucial role for understanding quench phenomena and therefore for improving quench detection. Yet, computational engineers struggle with the problem's complexity, which is the reason for the recent development of a hierarchical co-simulation framework for superconducting accelerator magnets called STEAM~\cite{Bortot_2018ab}. A major numerical challenge arises due to the multi-scale nature of both the magnet's geometry and the transient effects of a quench: While a superconducting accelerator magnet is over $10\,$m long, its diameter only measures a fraction of that~\cite{Bruning_2004aa}, and the cross-section contains many geometrical details (see Fig.~\ref{fig:benchmark}a). Additionally, quenches are very local effects leading to steep temperature fronts in longitudinal direction. Therefore, conventional three-dimensional (3D) finite-element (FE) simulations are computationally too expensive, as they would need very fine meshes to resolve the geometry and physical phenomena sufficiently, resulting into extremely huge systems of equations. This is further aggravated by the fact that it is a nonlinear magneto-thermal coupled problem, altogether leading to time-consuming simulations, e.g.~demonstrated in~\cite{Caspi_2003aa} and recently in~\cite{Troitino_2019aa}. 

In the context of STEAM~\cite{Bortot_2018ab}, this work presents an alternative approach by exploiting the geometrical translational invariance in longitudinal direction, which allows to use a quasi-3D (Q3D) ansatz. Herein, the transversal cross-section of the magnet is discretized with a two-dimensional (2D) FE method, while the longitudinal direction is resolved with a one-dimensional (1D) spectral-element (SE) method using orthogonal polynomials. In this way, the system's size can be significantly reduced. Similar approaches have been made e.g.~in~\cite{Jorks_2012aa} for a geometry with translation symmetry using higher-order FE methods, and~in~\cite{DAngelo_2017aa}-\cite{DAngelo_2019aa} for cylindrical structures using a Fourier expansion in azimuthal direction.

Here, the idea is illustrated for a thermal benchmark problem with linear material characteristics. The benchmark model represents a stack of Rutherford cables used for the coil in the superconducting magnet~\cite{Verweij_1995aa}. The method is validated against a 3D simulation using the commercial software COMSOL~\cite{COMSOL} and their computational efficiency is compared.

This work is structured as follows. First, the benchmark model is introduced and the problem setup is explained in Sec.~\ref{sec:benchmark}. In Sec.~\ref{sec:q3d}, the Q3D formulation for the heat conduction equation is elaborated. In Sec.~\ref{sec:results}, the simulation results are shown and discussed. Lastly, a conclusion of this work is given in Sec.~\ref{sec:conclusion}.

\section{Benchmark Model}\label{sec:benchmark}
\begin{table}[tbp]
	\caption{Benchmark model characteristics.}
	\begin{tabular}{|lc|lc|}
		\multicolumn{2}{|l|}{\textbf{Geometrical dimensions}} & \multicolumn{2}{|l|}{\textbf{Thermal conductivity} (W/m/K)} \\
		Cable width: & $1.5\,$mm & Homogenized cable: & 235.6 \\
		Cable height: & $15\,$mm & Glass fibre: & 0.01 \\
		Insulation thickness: & $100\,\upmu$m & & \\
		Total width: & $4.9\,$mm & \multicolumn{2}{|l|}{\textbf{Volumetric heat capacity} (J/m$^3$/K)} \\
		Total height: & $15.2\,$mm & Homogenized cable: & 314.1 \\
		Model length $\ell_z$: & $1\,$m & Glass fibre: & 750 \\
	\end{tabular}
\label{tab:benchmarksetup}
\end{table}
As benchmark model, a stack of three Rutherford cables wrapped by glass fibre insulation is considered, which is a component of the coil in superconducting magnets, see Fig.~\ref{fig:benchmark}. Each Rutherford cable contains a number of wires consisting of superconducting Nb$_3$Sn filaments embedded in a copper matrix. These wires are not modeled explicitly. Instead, the cables are considered to be solid and the material properties are homogenized. Table~\ref{tab:benchmarksetup} summarizes the geometrical dimensions and material characteristics of the benchmark model. Using a Cartesian coordinate system $(x,y,z)$, the $z$-direction is defined as the longitudinal direction such that the transversal cross-section lies in the $xy$-plane.

In the investigated scenario, the left cable quenches at a location $z_q = 0.33\,$m. The resulting heat excitation is described by a time-constant Gaussian function
\begin{equation}
	q(x,y,z) = \hat{q}\,\exp\left( - \frac{(z-z_q)^2}{\sigma^2} \right) \chi_q(x,y),
	\label{eq:excitation}
\end{equation}
where $\hat{q} = 10^6\,$W/m$^3$ is the excitation's amplitude, $\sigma = 0.05\,$m is the standard deviation and $\chi_q(x,y)$ is a characteristic function that is equal to one for coordinates $(x,y)$ in the left cable and zero otherwise. Furthermore, isothermal boundary conditions (BCs) with a fixed temperature $\theta_\mathrm{D} = 2\,$K are set at the back and front sides of the model, and adiabatic BCs are applied to the hull~$\Gamma_\mathrm{h}$. In the initial state, all cables are cooled down to $\theta_0 = 2\,$K. Then, the heat conduction in the benchmark model is described by the boundary-value problem
\begin{align}
	-\nabla \cdot \left( \lambda(\vec{r})\, \nabla \theta(\vec{r},t) \right) + \cv(\vec{r})\, \partial_t \theta(\vec{r},t) &= q(\vec{r},t), \label{eq:heatequation} \\
	\theta(x,y,z=0;t) = \theta(x,y,z=\ell_z;t) &= \theta_\mathrm{D}, \label{eq:bcdir} \\
	\left. -\lambda(\vec{r})\, \partial_n \theta(\vec{r},t) \right|_{\Gamma_\mathrm{h}} &= 0, \label{eq:bcneu} \\
	\theta(\vec{r},t=0) &= \theta_0. \label{eq:ic}  
\end{align}
Herein, $\lambda$ is the thermal conductivity in W/m/K, $\theta$ is the temperature in K, $\cv$ is the volumetric heat capacity in J/m$^3$/K, $\vec{r} = (x,y,z)$ is the spatial coordinate in m, and $t$ is the time in s. For the thermal steady-state solution, one has to solve~\eqref{eq:heatequation}-\eqref{eq:bcneu} with  $\partial_t \theta(\vec{r},t) = 0$.

\section{Quasi-3D Formulation}\label{sec:q3d}
\subsection{Spatial discretization and basis functions}
\begin{figure}[tbp]
	\centering
	\begin{subfigure}{3.5cm}
		\centering
		\includegraphics[height=4cm]{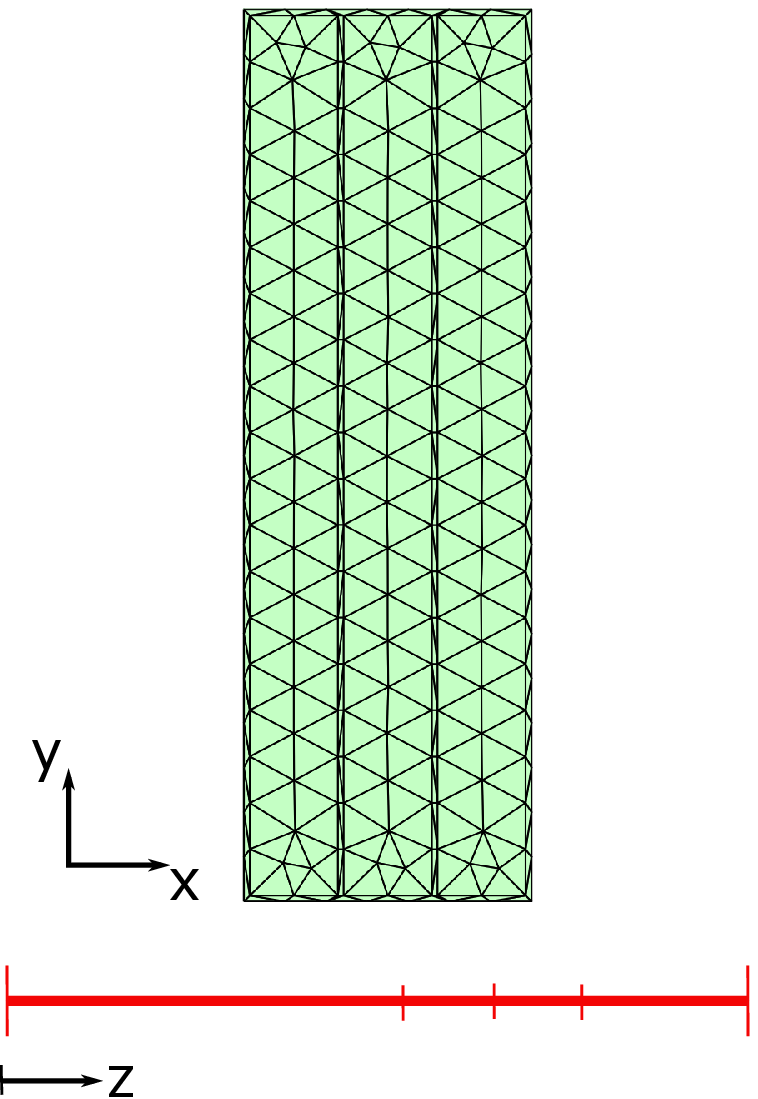}
		\caption{}
	\end{subfigure}
	\begin{subfigure}{5cm}
		\centering
		\includegraphics[height=4cm]{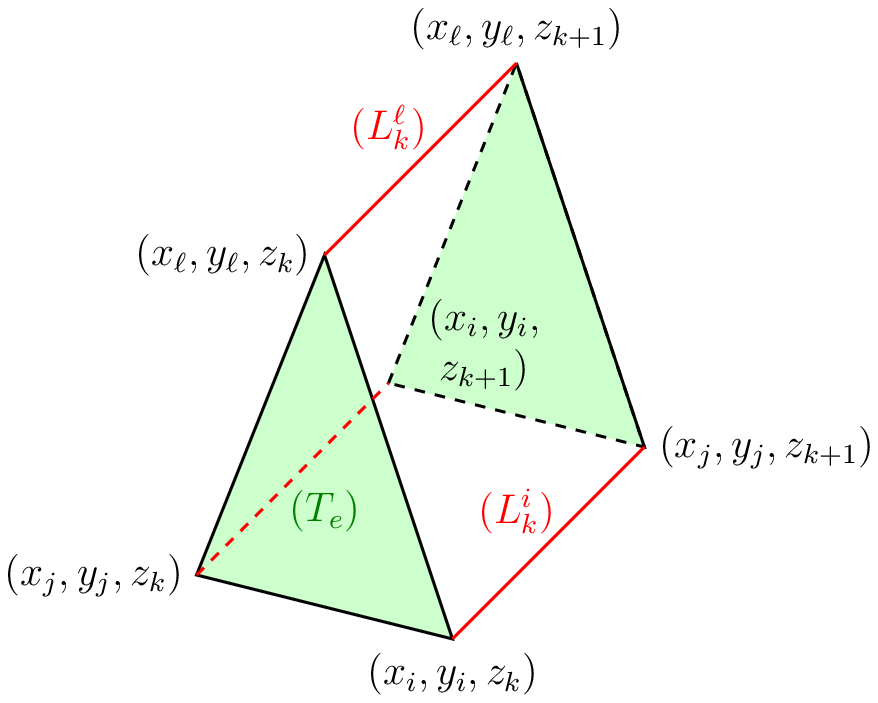}
		\caption{}
	\end{subfigure}
\caption{In (a), the triangular FE mesh in the transversal cross-section (green) as well as the 1D SE mesh in longitudinal direction (red) are shown. This discretization results into triangular prism elements (b).}
\label{fig:discretization}
\end{figure}
In the Q3D setting, the transversal cross-section is discretized with triangular FEs $T_e$ while the longitudinal direction is resolved with a 1D SE mesh, i.e.~with line elements $L_k$. As a result, the model's volume is assembled from triangular prism elements, see Fig.~\ref{fig:discretization}.
The temperature is discretized by
\begin{equation}
	\theta(x,y,z;t) \approx \sum\limits_{j=1}^J \sum\limits_{k=1}^K \sum\limits_{q=1}^{N+1} \widetilde{u}^{(k)}_{jq}(t) \, \shapefun(x,y,z),
\end{equation}
where $J$ is the number of FE nodes, $K$ is the number of SEs and $N$ is the maximal polynomial degree. The coefficients $\widetilde{u}^{(k)}_{jq}$ collect the degrees of freedom (DoFs). In contrast to the FE method, these coefficients do not entirely live in the \emph{physical space}, i.e.~they do not represent the temperature solution directly, but they form the expansion coefficients for the polynomials and thus represent something like the frequency of the solution. Therefore, it is said that they live in the \emph{spectral} or \emph{frequency space}~\cite{Shen_2011aa}. 
The shape functions $\shapefun(x,y,z)$ are expressed as a product
\begin{equation}
	\shapefun(x,y,z) = N_j(x,y) \, \phi^{(k)}_q (z),
\end{equation}
of $N_j(x,y)$ being the standard FE linear 2D nodal shape functions~\cite{Brenner_2008aa}, and $\phi^{(k)}_q(z)$ are chosen as modified Lobatto polynomials, which are defined on the orthogonality interval $I=[-1,1]$ globally as 
\begin{equation}
	\phi_q(\xi) = \left\lbrace 
	\begin{array}{ll}
		\frac{1-\xi}{2}, & q = 1, \\
		\frac{1-\xi^2}{4} \, \lobatto_{q-2}(\xi) & q = 2,\dots,N, \\
		\frac{1+\xi}{2} & q = N+1
	\end{array} \right.
	\label{eq:modifiedlobatto}
\end{equation}
with a mapped parameter $\xi(z) \in I$ and $\lobatto_q(\xi)$ being the Lobatto polynomial of $q$-th order~\cite{Pozridikis_2014aa}. The choice of this particular polynomials is motivated by the desire to achieve sparse matrices: The FE nodal shape functions are chosen to be \emph{nodal}, i.e.~each node only interacts with its neighboring nodes. Similarly, the SE polynomial shape functions are chosen to be \emph{modal}, i.e.~their interactions in the frequency space are reduced~\cite{Fakhar-Izadi_2015aa}. An exception are the boundary modes $\lbrace 1, N+1 \rbrace$, which are nodal and therefore represent the physical solution. In this way, although the SE matrices are not purely diagonal, the nodal boundary modes will make it easy to impose interface continuities and BCs. By using the Vandermonde matrix of the modified Lobatto polynomials~\cite{Pozridikis_2014aa}, it is possible to forward transform a physical solution into a spectral one, or to backward transform a spectral solution into a physical one.

\subsection{Discrete system}
\begin{figure}[tbp]
	\centering
	\includegraphics[width=1\columnwidth]{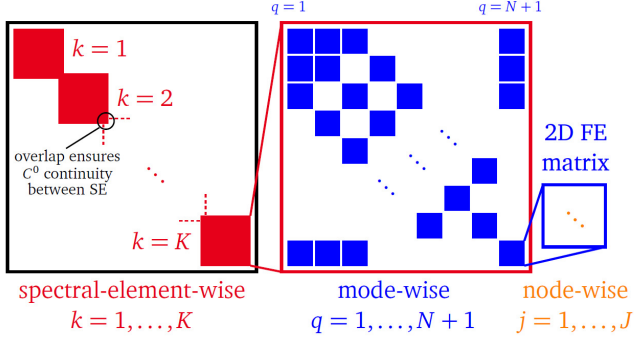}
	\caption{Tensor-product structure of the Q3D matrices: On the top level, there are block matrices for each SE which consist of a mode-wise block matrix pattern. These lowest level block matrices are the nodal 2D FE matrices.}
	\label{fig:matrixpattern}
\end{figure}
Applying the Ritz-Galerkin method to~\eqref{eq:heatequation}-\eqref{eq:bcneu} and choosing the test functions identically to the shape functions, $\testfun = \shapefun$, the system of equations
\begin{equation}
	\mathbf{K}^\mathrm{Q3D}_\lambda \, \widetilde{\mathbf{u}}(t) + \mathbf{M}^\mathrm{Q3D}_{\cv} \, \partial_t \widetilde{\mathbf{u}}(t) = \mathbf{q}^\mathrm{Q3D}
	\label{eq:system}
\end{equation}
arises. The Q3D matrices and vectors are constructed by the 2D Cartesian FE and 1D SE matrices and vectors by
\begin{align}
	\mathbf{K}^\mathrm{Q3D}_\lambda &= \mathbf{M}^\mathrm{SE} \otimes \mathbf{K}_\lambda^\mathrm{FE} + \mathbf{K}^\mathrm{SE} \otimes \mathbf{M}_\lambda^\mathrm{FE}, \\
	\mathbf{M}^\mathrm{Q3D}_{\cv} &= \mathbf{M}^\mathrm{SE} \otimes \mathbf{M}^\mathrm{FE}_{\cv}, \\
	\mathbf{q}^\mathrm{Q3D} &= \mathbf{q}^\mathrm{SE} \otimes \mathbf{q}^\mathrm{FE}.
\end{align}
Herein, $\otimes$ denotes the Kronecker tensor product. Fig.~\ref{fig:matrixpattern} sketches this tensor-product structure of the Q3D matrices. The FE quantities are found as
\begin{align}
	(\mathbf{K}^\mathrm{FE}_\lambda)_{ij} &= \int\limits_\Omega \lambda(x,y) \nabla N_j(x,y) \cdot \nabla N_i(x,y) \, \dd x \dd y, \\
	(\mathbf{M}^\mathrm{FE}_\alpha)_{ij} &= \int\limits_\Omega \alpha(x,y) N_j(x,y) N_i(x,y) \, \dd x \dd y, \\
	(\mathbf{q}^\mathrm{FE})_i &= \int\limits_\Omega q^t(x,y) N_i(x,y) \, \dd x \dd y
\end{align}
with $\Omega$ as the 2D cross-section and $\alpha$ standing for $\lambda$ and $\cv$, respectively. The element-wise SE quantities read
\begin{align}
	(\mathbf{K}^\mathrm{SE})_{pq}^{(k)} &= \int\limits_{I} \partial_z \phi_q^{(k)}(z) \, \partial_z \phi_p^{(k)}(z) \, \dd z, \label{eq:stiffness_se} \\
	(\mathbf{M}^\mathrm{SE})_{pq}^{(k)} &= \int\limits_{I} \phi_q^{(k)}(z) \, \phi_p^{(k)}(z) \, \dd z, \label{eq:mass_se} \\
	(\mathbf{q}^\mathrm{SE})_p^{(k)} &= \int\limits_{I} (\mathcal{I}_{N+1} \, q^\ell(z)) \, \phi_p^{(k)}(z) \, \dd z,
\end{align}
where $\mathcal{I}_{N+1}$ is the modified Lobatto interpolation operator. Furthermore, it is assumed that the excitation function can be separated according to $q(x,y,z) = q^t(x,y) q^\ell(z)$. The global SE matrices are built by ordering the element-wise SE matrices diagonally like illustrated in Fig.~\ref{fig:matrixpattern} on the left, where it is important to overlap the element-wise matrices' corners to ensure $C^0$-continuity across the SE interfaces~\cite{Fakhar-Izadi_2015aa}. The semi-discrete system~\eqref{eq:system} is discretized in time with an implicit backward Euler method. Eventually, the physical solution is obtained by a backward transform of the spectral solution.

\subsection{Adaptive spectral meshing}
\begin{figure}[tbp]
	\centering 
	\includegraphics[width=.8\columnwidth]{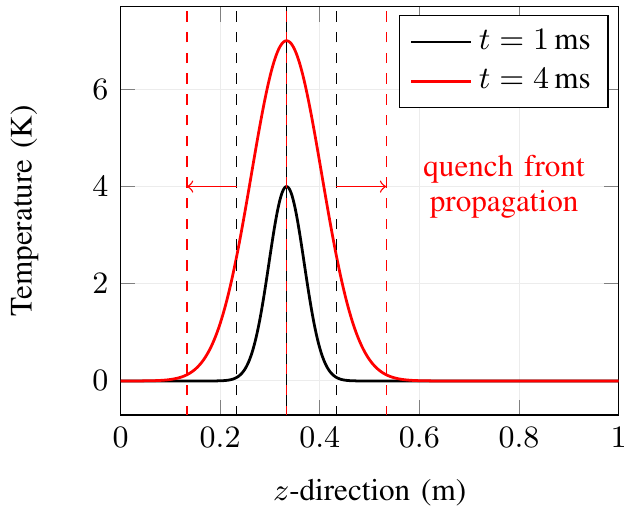}
	\caption{Temperature distribution in the quenched cable in $z$-direction for two different points in time. The dashed lines indicate the SE interfaces and their adaptation to the moving quench fronts.}
	\label{fig:adaptive}
\end{figure}
By choosing the SE interfaces appropriately, only a few polynomial orders suffice to resolve the quench phenomenon in $z$-direction. However, as the quench fronts move over time it becomes necessary to adapt the SE interfaces to them, because otherwise more polynomials would be required to represent the temperature distribution well, see Fig.~\ref{fig:adaptive}. Even worse, steep temperature fronts within a SE may produce the \emph{Gibbs phenomenon} or \emph{spectral ringing}~\cite{Boyd_2001aa}. Therefore, an adapted spectral meshing of the longitudinal direction is mandatory to achieve accurate results with affordable computational costs. 

This is done by regularly checking the solution during the simulation and shifting the old SE interfaces (Fig.~\ref{fig:adaptive}, dashed black lines) according to the quench fronts (red dashed lines) when necessary. The physical solution is then forward transformed onto the new spectral mesh, and the matrices are reassembled. In this process, the structure and size of the matrices are maintained, as only the mapping coefficients need to be recalculated.

\section{Simulation Results}\label{sec:results}
\begin{figure}[tbp]
	\centering
	\includegraphics[width=.8\columnwidth]{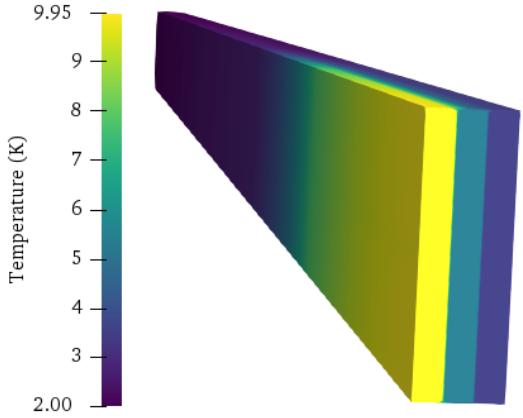}
	\caption{Temperature distribution in the benchmark model after $10\,$ms. The geometry has been sliced at the hot-spot level $z=z_q$.}
	\label{fig:paraview}
\end{figure}
The benchmark problem described in Sec.~\ref{sec:benchmark} is computed with the proposed Q3D method and validated against a conventional 3D simulation using COMSOL~\cite{COMSOL}. The temperature distribution after $10\,$ms is visualized in Fig.~\ref{fig:paraview}, where the model has been sliced at $z=z_q$ enabling to see the transversal temperature distribution. As a consequence of the glass fibre insulation, the heat exchange is faster in longitudinal than in transversal direction. As a result, there are sharp temperature gradients from one cable to another, while the temperature varies smoothly in longitudinal direction.

\begin{figure}[tbp]
	\centering 
	\includegraphics[width=.9\columnwidth]{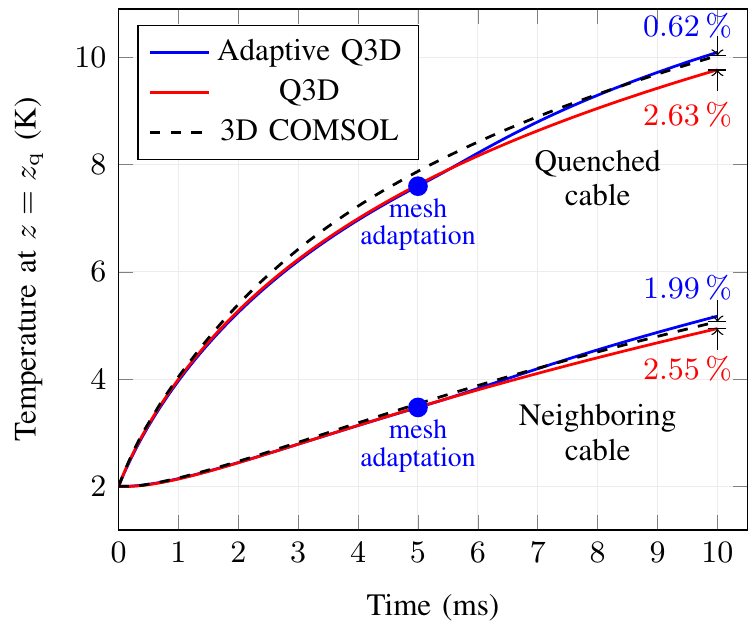}
	\caption{Temperature curve over time at $z=z_q$ for the quenched cable and its neighboring cable. The reference solution by COMSOL is depicted as black dashed line. The red curve shows the Q3D solution without making a spectral mesh adaption, while the blue curve depicts the Q3D solution after adapting the spectral mesh after half-time of the simulation (indicated with a blue dot).}
	\label{fig:validation}
\end{figure}
Fig.~\ref{fig:validation} shows the resulting temperature curves for the hot-spot point at $z=z_q$ in the quenched cable as well as in the neighboring cable. Here, the Q3D solution (red) aligns well with the reference solution (black) leading to relative differences below $3\,\%$ at the end of the simulation time. When adapting the spectral mesh after half-time of the simulation to the new quench front locations, the Q3D solution (blue) is slightly improved and delivers relative differences below $2\,\%$.

\begin{figure}[tbp]
	\centering 
	\includegraphics[width=.9\columnwidth]{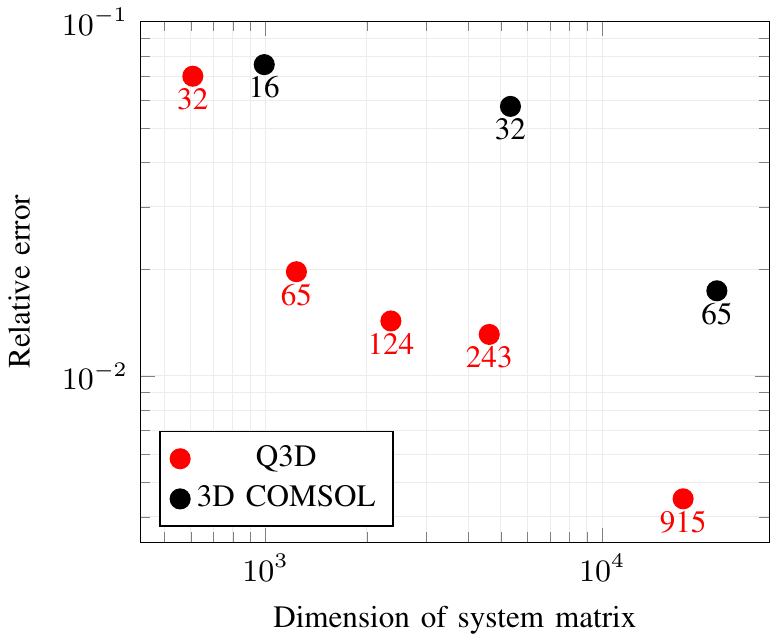}
	\caption{The dimension of the system matrix in the Q3D (red) and 3D COMSOL (black) case are plotted against the relative error w.r.t.~a fine reference solution computed by COMSOL. The numbers below the marks denote the number of triangular elements resolving the transversal cross-section. The reference solution itself is resolved with 252 triangles.}
	\label{fig:efficiency}
\end{figure}
To compare the spatial accuracy of the Q3D and 3D method, steady-state simulations are carried out to exclude time stepping errors.
The computational efficiency is visualized in Fig.~\ref{fig:efficiency}, where the dimension of the system matrix is plotted against the relative error of the hot-spot temperature w.r.t.~a fine COMSOL reference solution with 252 triangular elements in the cross-section. Clearly, the Q3D method yields more accurate results with less computational effort than the 3D simulation, e.g.~for a cross-section resolution with 65 triangles, the 3D simulation requires a system matrix of size $21898\times21898$, while the Q3D method achieves the same transversal resolution and a similar error with a $98\,\%$ smaller system matrix of size $2405\times2405$. It must be noted that the benchmark model has a length of $\ell_z = 1\,$m, while the real geometry of a superconducting accelerator magnet is over $10\,$m long. Consequently, it is expected that the efficiency gain obtained by the Q3D approach will be much more pronounced for the real magnet geometry.

\section{Conclusion}\label{sec:conclusion}
To deal with the numerical challenges arising in the simulation of quench propagation in superconducting magnets, a Q3D FE-SE method has been presented as an alternative to the conventional 3D FE method. As a proof of concept, the heat conduction in a Rutherford cable stack has been computed and successfully validated against a 3D FE simulation using COMSOL. In terms of computational efficiency, the Q3D approach has proved to be superior to the conventional 3D FE method. Therefore, the presented Q3D method is a promising and efficient numerical tool to accurately simulate quench propagation phenomena in superconducting accelerator magnets.

\end{document}